\documentclass[aps,pra,showpacs,twocolumn,superscriptaddress]{revtex4-1}
\usepackage[latin9]{inputenc}
\usepackage{amsmath}
\usepackage{amssymb}
\usepackage{graphicx}
\usepackage[unicode=true,
 bookmarks=false,
 breaklinks=false,pdfborder={0 0 1},backref=section,colorlinks=false]
 {hyperref}
\hypersetup{
 dvipdfm,colorlinks}
\usepackage{breakurl}

\makeatletter

\newcommand*\LyXThinSpace{\,\hspace{0pt}}
\providecommand{\tabularnewline}{\\}

\makeatother

\begin{document}

\title{Multiqubit Greenberger-Horne-Zeilinger state generated by synthetic
magnetic field in circuit QED}

\author{Wei Feng}

\affiliation{Beijing Computational Science Research Center, Beijing 100193, China}

\affiliation{Texas A\&M University, College Station, TX 77843, USA}

\affiliation{Department of Physics, Zhejiang University, Hangzhou 310027, China}

\author{Da-Wei Wang }

\affiliation{Texas A\&M University, College Station, TX 77843, USA}

\author{Han Cai}

\affiliation{Texas A\&M University, College Station, TX 77843, USA}

\author{Shi-Yao Zhu}

\affiliation{Beijing Computational Science Research Center, Beijing 100193, China}

\affiliation{Department of Physics, Zhejiang University, Hangzhou 310027, China}

\date{\today }
\begin{abstract}
We propose a scheme to generate Greenberger-Horne-Zeilinger (GHZ)
state for \textit{N} superconducting qubits in a circuit QED system.
By sinusoidally modulating the qubit-qubit coupling, a synthetic magnetic
field has been made which broken the time-reversal symmetry of the
system. Directional rotation of qubit excitation can be realized in
a three-qubit loop under the artificial magnetic field. Based on the
special quality that the rotation of qubit excitation has different
direction for single- and double-excitation loops, we can generate
three-qubit GHZ state and extend this preparation method to arbitrary
multiqubit GHZ state. Our analysis also shows that the scheme is robust
to various operation errors and environmental noise. 
\end{abstract}
\maketitle

\section{introduction}

Entanglement, which lets the measurement of one particle instantaneously
determine the quantum state of a partner particle, is a salient nonclassical
feature of quantum physics \cite{EPR}. Besides the use in testing
of fundamental quantum theories \cite{Bell,Pan-Nature,Bell test},
entangled state plays a fundamental role in quantum technology applications
such as quantum computation, quantum communication \cite{nielsen},
quantum simulation \cite{Quantum simulation} and quantum-enhanced
precision measurements \cite{Measurements}. As the maximally entangled
state of three or more particles, Greenberger-Horne-Zeilinger (GHZ)
states \cite{GHZ} represent a paradigmatic multipartite entangled
states that are, in particular, useful for error correction in quantum
computing and quantum secret sharing \cite{Use of GHZ}. Many efforts
have been devoted to generation of GHZ states in different qubit systems,
including photons \cite{photon-threeGHZ,photon-three-GHZ-realize,10-photon GHZ},
atoms in cavity QED \cite{atom-GHZ}, trapped ions \cite{Molmer,GHZ14},
Bose-Einstein condensed atoms \cite{BEC-GHZ,BEC-GHZ2}, nuclear spins
in nitrogen-vacancy (NV) defect center \cite{NV-center}, and mechanical
resonators in optomechanical system \cite{Optomechanical GHZ}. 

The circuit QED, in which superconducting circuits based on Josephson
junctions serve as artificial atoms, has many advantages such as tunability,
flexibility, and fabricating on solid-state chip with standard lithographical
technologies \cite{CQED-Review,CQED-RMP,CQED-review2,CQED-science}.
It is a promising candidate among many architectures for realization
of quantum computation and quantum information protocols. Moreover,
in contrast to natural atoms, superconducting circuits have strong
coupling with electromagnetic fields and can be designed with special
characteristics. These merits made it to be a ideal artificial system
in which many phenomenons of quantum optics can be studied in a regime
which was difficult to achieve with natural atoms \cite{Feng Wei}.
Recently, experiments have great progress on the subjects related
to circuit QED, such as generation of arbitrary quantum states of
a single superconducting resonator \cite{Martinis-Nature-2009}, realization
of tunable qubit-qubit coupling \cite{Martinis-PRL-2014}, and arranging
qubits in the form of a lattice \cite{Circuit QED Lattices}.

Under the drive of huge potential application, many different protocols
have been proposed to generate GHZ states in circuit QED setups \cite{WangYinDan-circuit_QED,WangYinDan-circuit_QED2,cqed-GHZ,cqed-GHZ1,cqed-GHZ2,cqed-GHZ3,cqed-GHZ4,cqed-GHZ5}.
Some of them are based on measurement, i.e., if a special measurement
has a special result, the system is known to be in a GHZ state after
the measurement \cite{cqed-GHZ1,cqed-GHZ3,cqed-GHZ5}. These methods
are of a probabilistic nature or need cumbersome feedback control.
Some others are one-step schemes using a Mølmer-Sørensen approach
\cite{WangYinDan-circuit_QED,WangYinDan-circuit_QED2}, i.e., a effective
Hamiltonian of the type $\hat{J}_{x}^{2}$ will produce a GHZ state
after a definite duration \cite{Molmer}. These methods are deterministic
and efficient, but it is hard to realize homogeneous coupling among
qubits, especially for large-number GHZ state generation. To date,
10-qubit and 14-qubit GHZ states have been successfully generated
in experiments for photons \cite{10-photon GHZ} and trapped ions
\cite{GHZ14}, respectively. However, in recent report, the number
of experimentally realized \textbf{\textit{N}}-qubit GHZ state for
superconducting circuit is just three and five \cite{three-GHZ1,three-GHZ2,Five-GHZ}.
Therefore, new schemes which combine recent experimental progress
on circuit QED are highly needed to generate multiqubit GHZ state
for superconducting circuit. 

In the present paper, we introduce a new mechanism to generate GHZ
states in a systematic way. The method is inspired by a recent experimental
work in which the chiral ground-state currents are detected under
a synthetic magnetic field \cite{naturePhysics_Martinis}. Using a
recent technology of sinusoidally modulating the qubit-qubit coupling\cite{Martinis-PRL-2014},
an effective resonance hopping between qubits with a complex hopping
amplitude can be realized. Under this interaction, a synthetic magnetic
field has been made which broken the time-reversal symmetry of the
system. For three-qubit loop of superconducting circuits, there is
a novel character: when the hopping amplitudes are pure imaginary
numbers, the qubit excitation will have directional rotation and the
directions are different for single- and two-excitation cases. Owing
to this character, we can generate three-qubit GHZ state and extend
this preparation method to arbitrary multiqubit GHZ state. Our analysis
also shows that the scheme is robust to various operation errors and
environmental noise. 

This paper is organized as following. In Sec. II, we derive on effective
Hamiltonian with qubit-qubit interaction amplitude is a complex number,
and show the directional rotation of excitation in three-qubit loop.
In Sec. III, we describe the protocol for generating GHZ states in
our system. In Sec. IV, we discuss experimental feasibility considering
the environmental decoherence and operation errors. Finally, we make
a conclusion in Sec. \mbox{V}. 

\section{effective hamiltonian and excitation circulation for three-qubit
loop}

We consider a system consisting of three qubits where tunable coupling
is exist between each two of them. The Hamiltonian of the system is 

\begin{equation}
H\left(t\right)=\hbar\sum_{i=1}^{3}\frac{\omega_{i}}{2}\sigma_{i}^{z}+\hbar\sum_{i\neq j}g_{ij}\left(t\right)\left(\sigma_{i}^{+}\sigma_{j}^{-}+\sigma_{j}^{+}\sigma_{i}^{-}\right),
\end{equation}
where $\omega_{i}$ is the frequency of qubit $Q_{i}$, $g_{ij}$
is the strength of the inter-qubit coupling between qubits $Q_{i}$
and $Q_{j}$. The Pauli matrices $\sigma^{z}=\left|1\right\rangle \left\langle 1\right|-\left|0\right\rangle \left\langle 0\right|$,
and $\sigma^{+}=\left|1\right\rangle \left\langle 0\right|$ ($\sigma^{-}=\left|0\right\rangle \left\langle 1\right|$)
is the raising (lowering) operators. $\left|0\right\rangle $ and
$\left|1\right\rangle $ denote the two energy level of the qubit.
We modulate the coupling strength according to $g_{ij}\left(t\right)=g_{0}\cos\left(\Delta_{ij}t+\phi_{ij}\right)$,
and set $\Delta_{ij}=\omega_{i}-\omega_{j}$. Under the condition
that $\left|g_{0}\right|\ll\left|\Delta_{ij}\right|$, we can do the
rotating-wave approximation, and the effective Hamiltonian of the
system in the interaction picture is 

\begin{equation}
H_{I}=\hbar\sum_{i\neq j}\frac{g_{0}}{2}\left(e^{i\phi_{ij}}\sigma_{i}^{+}\sigma_{j}^{-}+e^{-i\phi_{ij}}\sigma_{j}^{+}\sigma_{i}^{-}\right).\label{eq:effective Hamiltonian}
\end{equation}
\begin{figure}
\begin{centering}
\includegraphics[width=9cm]{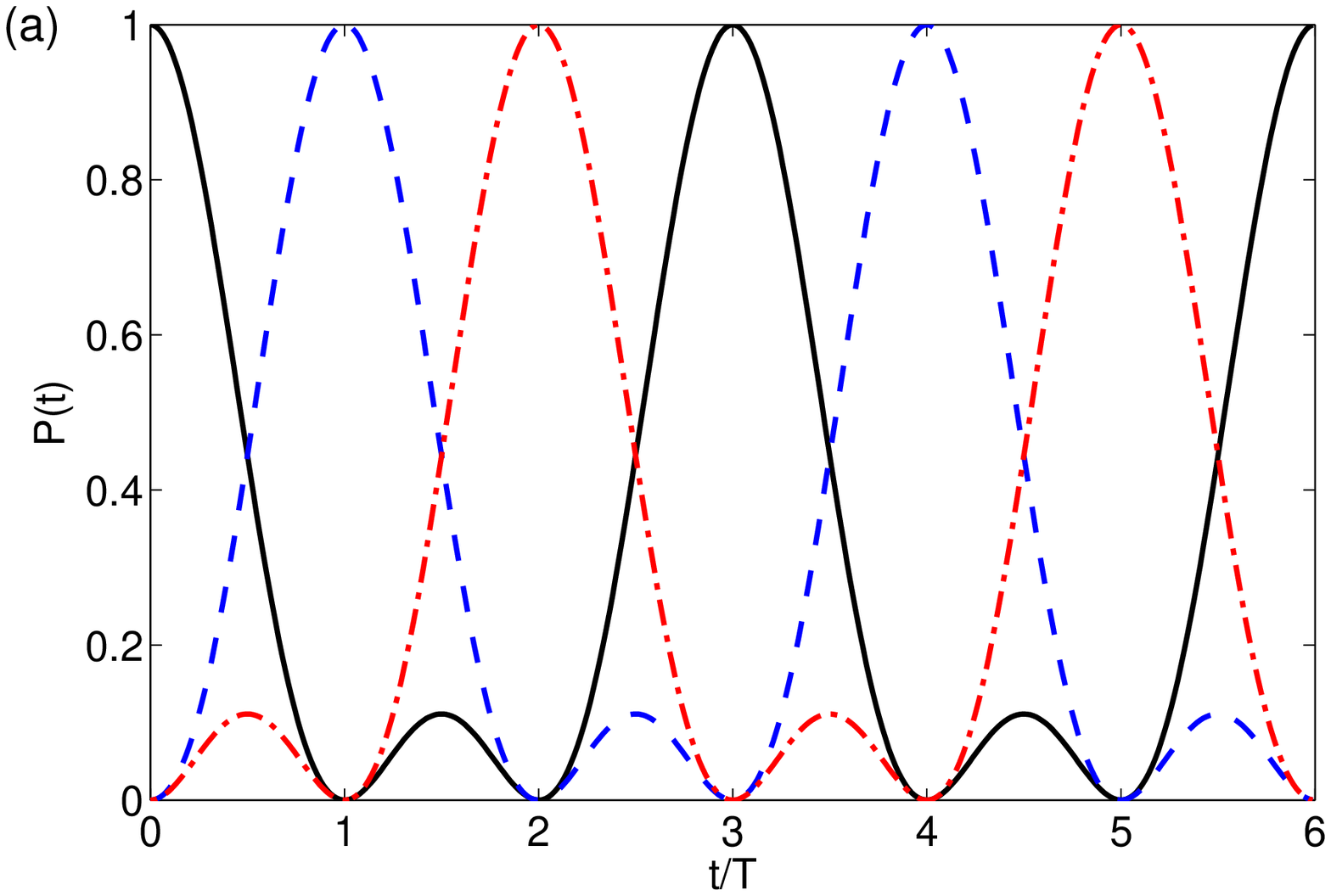}
\par\end{centering}
\begin{centering}
\includegraphics[width=9cm]{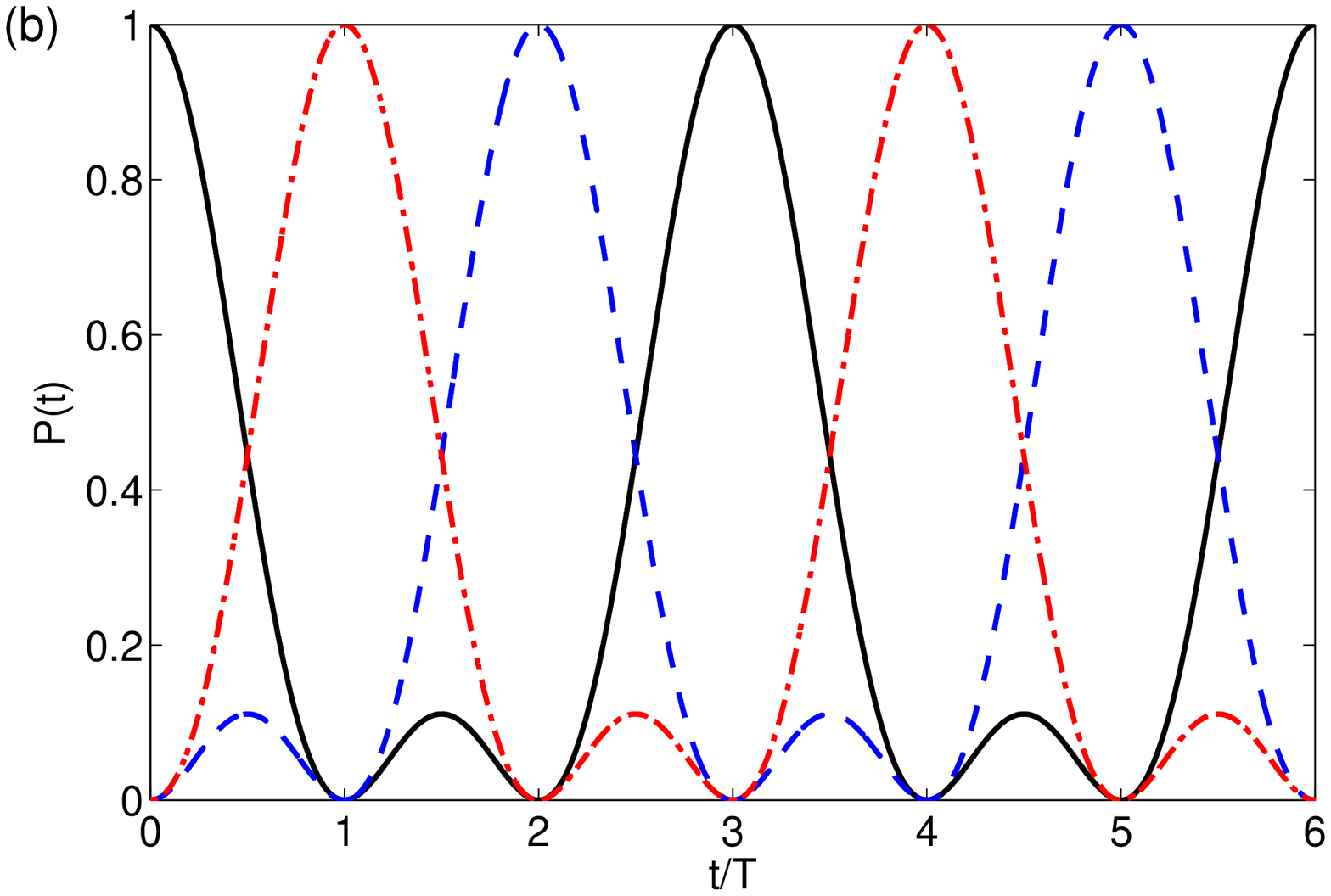}
\par\end{centering}
\caption{(Color online) The time evolution of the state occupation probabilities
for the three-qubit system with $\phi_{ij}=\pi/2$. (a) One-excited
case with initial state $|100\rangle$. The excitation from qubit
$Q_{1}$ (black solid line) is swapped to qubit $Q_{2}$ (blue dash
line) at time $T$ and then to qubit $Q_{3}$ (red dash dot line)
at time $2T$, and then repeat this circle. (b) Two-excited case with
initial state $|011\rangle$, the population of $|011\rangle$ (black
solid line) transfer to $|110\rangle$ (red dash dot line) at time
$T$ and then to $|101\rangle$ (blue dash line) at time $2T$, and
then repeat this circle.\label{fig:population}}
\end{figure}

Now we consider the cases that inter-qubit coupling amplitudes are
pure imaginary numbers, i.e., set $\phi_{ij}=\pi/2$. The Hamiltonian
in Eq. (\ref{eq:effective Hamiltonian}) commutes with $\sum_{i}\sigma_{i}^{z}$
and the number of excitation is conserved. We first investigate the
single-excitation dynamics in the subspace expanded by $|100\rangle$,
$|010\rangle$ and $|001\rangle$, in which the Hamiltonian has the
following matrix representation 
\begin{equation}
M_{1}=\hbar\frac{g_{0}}{2}\left(\begin{array}{ccc}
0 & i & -i\\
-i & 0 & i\\
i & -i & 0
\end{array}\right).
\end{equation}
The eigen frequencies are $\lambda_{1}=0$, $\lambda_{2}=\frac{\sqrt{3}}{2}g_{0}$
and $\lambda_{3}=-\frac{\sqrt{3}}{2}g_{0}$. The corresponding eigenstates
are 
\begin{equation}
\begin{aligned} & |\psi_{1}\rangle=\frac{1}{\sqrt{3}}\left(|100\rangle+|010\rangle+|001\rangle\right),\\
 & |\psi_{2}\rangle=\frac{1}{\sqrt{3}}\left(|100\rangle+e^{i2\pi/3}|010\rangle+e^{i4\pi/3}|001\rangle\right),\\
 & |\psi_{3}\rangle=\frac{1}{\sqrt{3}}\left(|100\rangle+e^{i4\pi/3}|010\rangle+e^{i2\pi/3}|001\rangle\right).
\end{aligned}
\end{equation}
If we initially prepare the state $|\Psi(0)\rangle=|100\rangle=\left(|\psi_{1}\rangle+|\psi_{2}\rangle+|\psi_{3}\rangle\right)/\sqrt{3}$,
the evolution of the wavefunction is 
\begin{equation}
\begin{aligned}|\Psi\left(t\right)\rangle= & \frac{1}{\sqrt{3}}\sum\limits _{j=1}^{3}e^{-i\lambda_{j}t}|\psi_{j}\rangle\\
= & \frac{1}{3}\left[\left(1+2\cos\left(\frac{\sqrt{3}}{2}g_{0}t\right)\right)|100\rangle\right.\\
 & +\left(1+2\cos\left(\frac{\sqrt{3}}{2}g_{0}t-2\pi/3\right)\right)|010\rangle\\
 & +\left.\left(1+2\cos\left(\frac{\sqrt{3}}{2}g_{0}t+2\pi/3\right)\right)|001\rangle\right].
\end{aligned}
\label{wf}
\end{equation}
It is clear that at time $t=T\equiv4\pi/(3\sqrt{3}g_{0})$, $|\Psi\left(T\right)\rangle=|010\rangle$,
and at time $t=2T$, $|\Psi\left(2T\right)\rangle=|001\rangle$, as
shown in Fig.\ref{fig:population} (a). 

\begin{figure}
\begin{centering}
\includegraphics[width=9cm]{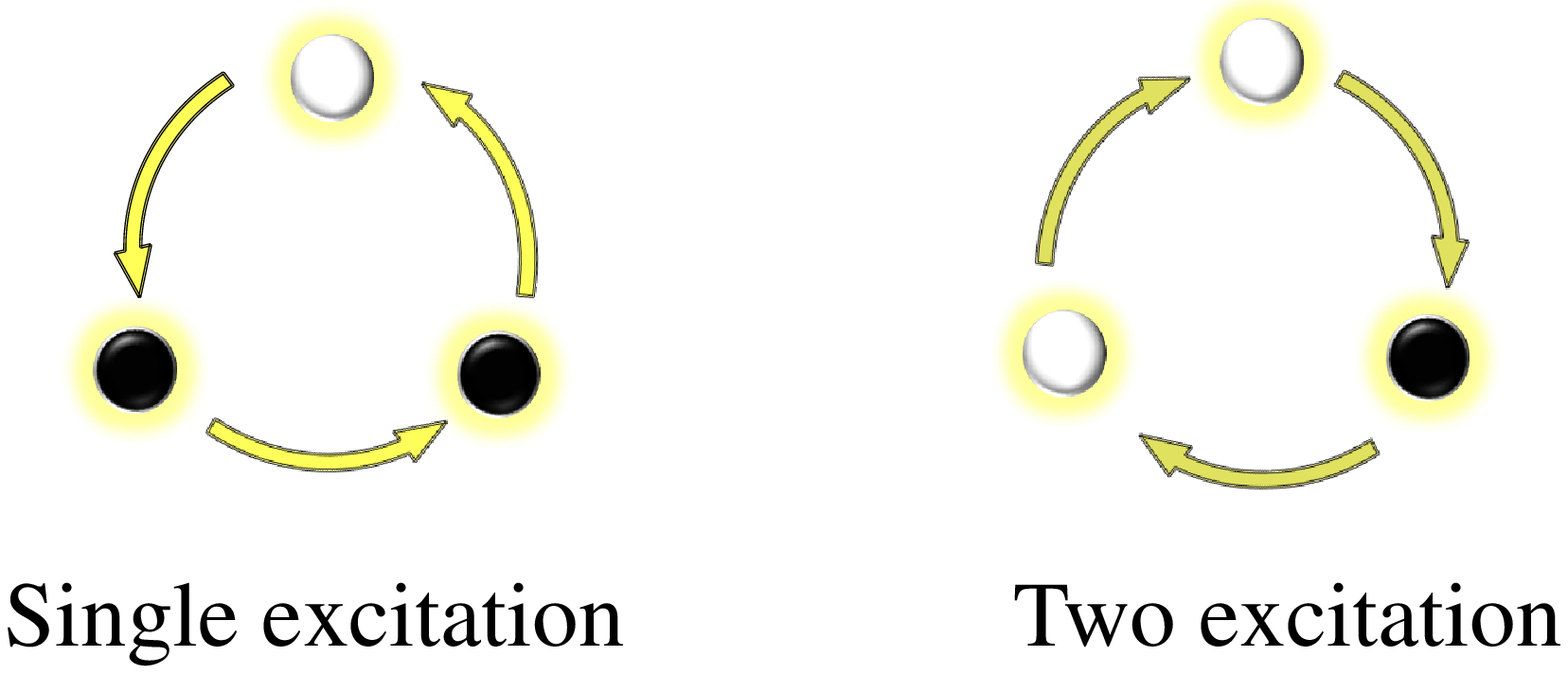}
\par\end{centering}
\caption{Excitation circulation for single- and two-excitation cases in three-qubit
loop. The white dot stand for qubit in the excited state and the black
dot stand for qubit in the ground state. While the single excitation
rotate in the anticlockwise direction, the two excitations rotate
in the reverse direction. \label{fig:Excitation-circulation-for}}

\end{figure}

Next we consider the subspace containing two excitations, $|011\rangle$,
$|101\rangle$ and $|110\rangle$, the reversed states of the three
states in the single excitation subspace. The corresponding matrix
of the Hamiltonian is 
\begin{equation}
M_{2}=-M_{1}.
\end{equation}
The overall minus sign indicates a excitation rotation in the opposite
direction. The evolution of an initial state $|011\rangle$ can be
obtained by changing $g_{0}$ to $-g_{0}$ and reversing the states
in Eq. (\ref{wf}). The time evolution of the state occupation probabilities
is plotted in Fig. \ref{fig:population} (b). For clearly see the
difference of the time evolution of above two cases, we also plot
a schematic figure for the single-excitation and two-excitation circulation,
see Fig. \ref{fig:Excitation-circulation-for} .

The surprising results that the single-excitation and two-excitation
cases have different circulation directions can be understood as following.
The state $|011\rangle\rightarrow|100\rangle$ can be seen as simply
reverse the definition of $\left|0\right\rangle $ and $\left|1\right\rangle $.
To express the Hamiltonian in this reversed new basis, we need to
switch $\sigma_{j}^{+}\rightarrow\sigma_{j}^{-}$ and $\sigma_{j}^{-}\rightarrow\sigma_{j}^{+}$,
which results in $H\rightarrow-H$. The state will evolve backward
in time equivalently. Actually, the condition to produce this different
circulations for single- and two-excitation cases, i.e., $\phi_{ij}=\pi/2$,
can be loosen to $\Phi=\phi_{12}+\phi_{23}+\phi_{31}=\pm\pi/2$. The
total phase $\Phi$ can be seen as an effective magnetic flux. The
restrictive phase of oscillating coupling introduce a effective magnetic
field, which produce the circulation of qubit excitation.

\section{protocol for generating ghz states}

\begin{figure}
\begin{centering}
\includegraphics[width=9cm]{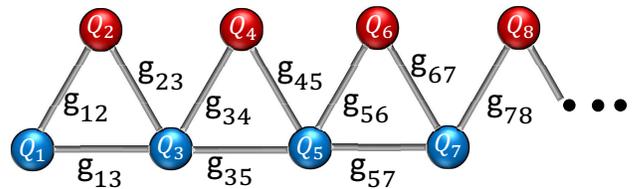}
\par\end{centering}
\caption{Schematic diagram of circuit QED lattice. The coupling between any
two qubits is tunable $g_{ij}\left(t\right)=g_{0}\cos\left(\Delta_{ij}t+\phi_{ij}\right),$
with $\Delta_{ij}$ equal to the frequency difference between $Q_{i}$
and $Q_{j}$.The simplest operation for realizing our protocol is
use $\omega_{1}=\omega_{3}=\omega_{5}=\omega_{7}\cdots,$ $\omega_{2}=\omega_{4}=\omega_{6}=\omega_{8}\cdots$,
$\Delta_{13}=\Delta_{35}=\Delta_{57}=\cdots=0,$ $\Delta_{12}=\Delta_{23}=\Delta_{34}=\Delta_{45}=\cdots=\omega_{1}-\omega_{2},$
$\phi_{12}=\phi_{34}=\phi_{56}=\cdots=\pi/2$, and $\phi_{23}=\phi_{45}=\phi_{67}=\cdots=0$.
\label{fig:Schematic-diagram}}
\end{figure}

The directional circulation of excitation allow us to produce GHZ
state by turning on the interaction for a definite duration and manipulating
a sequence of electromagnetic pulses. Firstly, we prepare three-qubit
GHZ state, and then extend it to produce large-number GHZ states.
The devise employed here is a superconducting circuit lattice \cite{Circuit QED Lattices,naturePhysics_Martinis}.
A schematic diagram is shown in Fig. \ref{fig:Schematic-diagram}.
With mature nanofabrication techniques, the circuit QED can be assemble
as design. The most important experimental technology for realizing
our scheme is tunable coupling between qubits, which is reported in
reference \cite{Martinis-PRL-2014}.

\begin{figure}
\begin{centering}
\includegraphics[width=9cm]{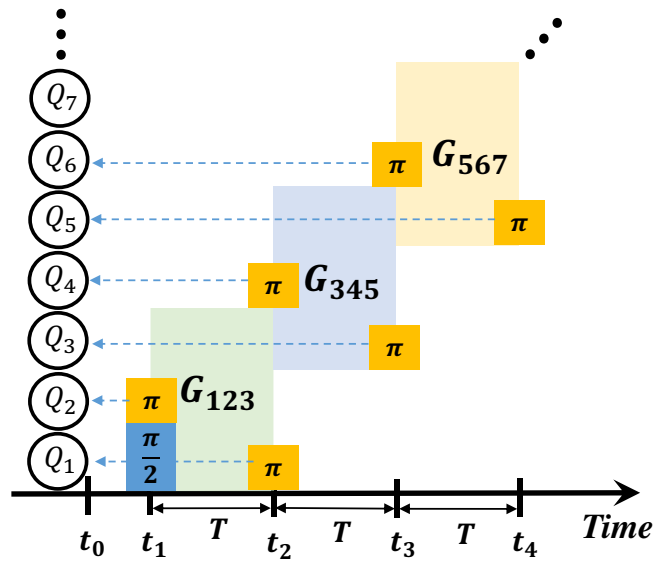}
\par\end{centering}
\caption{Sequence of operations. The blue box stand for a $\pi/2$ pulse and
the yellow box stand for a $\pi$ pulse. They act on the qubits that
arrows point to at the determinate points in time $t_{i}$. The characteristic
time for excitation swapping is $T=4\pi/(3\sqrt{3}g_{0})$. The shaded
area in color label with $G_{123}$($G_{345}$, $G_{567}$) mean turning
on the couplings in the corresponding three-qubit loop.\label{fig:Sequence-of-operations.}}
\end{figure}

\begin{table*}[t]
\begin{centering}
\begin{tabular}{|c|c|}
\hline 
time & states transfer under $\pi/2$ and $\pi$ pulse\tabularnewline
\hline 
\hline 
$t_{1}$ & $|0000000\cdots0\rangle$$\rightarrow$$\frac{1}{\sqrt{2}}\left(|0\rangle+|1\rangle\right)_{1}\otimes|100000\cdots0\rangle$\tabularnewline
\hline 
$t_{2}$ & $\frac{1}{\sqrt{2}}\left(|100\rangle+|011\rangle\right)_{123}\otimes|0\cdots00\rangle\rightarrow\frac{1}{\sqrt{2}}\left(|000\rangle+|111\rangle\right)_{123}\otimes|1\cdots00\rangle$\tabularnewline
\hline 
$t_{3}$ & $\frac{1}{\sqrt{2}}\left(|00100\rangle+|11011\rangle\right)_{12345}\otimes|0\cdots00\rangle\rightarrow\frac{1}{\sqrt{2}}\left(|00000\rangle+|11111\rangle\right)_{12345}\otimes|1\cdots00\rangle$\tabularnewline
\hline 
$t_{4}$ & $\frac{1}{\sqrt{2}}\left(|0000100\rangle+|1111011\rangle\right)_{1234567}\otimes|0\cdots00\rangle\rightarrow\frac{1}{\sqrt{2}}\left(|0000000\rangle+|1111111\rangle\right)_{1234567}\otimes|0\cdots00\rangle$\tabularnewline
\hline 
\end{tabular}
\par\end{centering}
\begin{widetext}
\caption{State evolution, basic steps of generation GHZ states. $t_{1}$: prepare
the first qubits in a superposition of 0 and 1 state, flip the second
qubits. $t_{1}\rightarrow t_{2}$: introduce an interaction $G_{123}$
among first three qubits, which leads to opposite rotations of the
two states as shown in Eq.(\ref{eq1}). $t_{2}$, send in $\pi$ pulse
to flip the first and fourth qubits. $t_{2}\rightarrow t_{3}$: introduce
the same interaction as in $t_{1}\rightarrow t_{2}$ with $G_{345}$.
$t_{3}$, send in $\pi$ pulse to flip the third and sixth qubits.
Repeat steps to realize GHZ state for larger number of qubits, each
time adding two qubits to the GHZ chain. For a GHZ state with $2n+1$
qubits, we need in total $2n$ $\pi$ pulses and one $\pi/2$ pulse.
\label{tab:State-evolution,-basic}}
\end{widetext}

\end{table*}

The sequence of operations is shown in Fig. \ref{fig:Sequence-of-operations.}.
We start from ground state $|000\cdots00\rangle$. At time $t_{1}$,
we send a $\pi/2$ pulse to $Q_{1}$ and a $\pi$ pulse to $Q_{2}$.
Qubit $Q_{1}$ is been prepared in a superposition state $\frac{1}{\sqrt{2}}\left(|0\rangle+|1\rangle\right)$
and $Q_{2}$ is been flipped to $|1\rangle$ state. The state of system
transfer to $\frac{1}{\sqrt{2}}\left(|0\rangle+|1\rangle\right)_{1}\otimes|100\cdots00\rangle$.
The following step is crucial. We switch on the couplings among the
first three qubits (denoted by $G_{123}$), as demonstrated in above
section, the three-qubit loop undergo opposite rotations for $\left|010\right\rangle $
and $\left|110\right\rangle $ states. 

\begin{equation}
\begin{aligned} & |010\rangle\rightarrow|100\rangle\rightarrow|001\rangle\rightarrow|010\rangle,\\
 & |110\rangle\rightarrow|011\rangle\rightarrow|101\rangle\rightarrow|110\rangle.
\end{aligned}
\label{eq1}
\end{equation}

After a definite time $T=4\pi/(3\sqrt{3}g_{0})$, turn off the interactions
$G_{123}$, the system evolve to $\frac{1}{\sqrt{2}}\left(|100\rangle+|011\rangle\right)_{123}\otimes|0\cdots00\rangle$
state. Then, we send a $\pi$ pulse to flip the first qubit. The first
three qubits are in a GHZ state. Next we continue our production of
multiqubit GHZ state: send a $\pi$ pulse to the fourth qubit and
switch on the same special couplings among the third, fourth and fifth
qubits (denoted by $G_{345}$). After these three-qubit states undergo
the process $|010\rangle\rightarrow|100\rangle$ and $|110\rangle\rightarrow|011\rangle$
in Eq. (\ref{eq1}) for the two component of the superposition state,
respectively, we obtain the state $\frac{1}{\sqrt{2}}\left(|00100\rangle+|11011\rangle\right)_{12345}$.
Then we send a $\pi$ pulse to the third qubit, the first five qubits
are prepared in a 5-qubit GHZ state, as shown in Tab. \ref{tab:State-evolution,-basic}.
We can repeat the process to generate the GHZ state involving more
qubits.

Above scheme is only for odd-number GHZ states. For even-number GHZ
states, we can use same method base on two-qubit Bell state $\frac{1}{\sqrt{2}}\left(|00\rangle+|11\rangle\right)_{12}$.
The entangled Bell states for superconducting qubit have been validly
studied and realized in several experiments \cite{Bell-state1,Bell-state2,Bell-staet3,Bell-state4}.
Starting in the ground state, a $\pi/2$ pulse is applied to qubit
$Q_{1}$ to create the superposition $\frac{1}{\sqrt{2}}\left(|0\rangle+|1\rangle\right)_{1}\otimes|0\cdots00\rangle$.
Next, a CNOT gate is applied to flip qubit $Q_{2}$ conditioned on
qubit $Q_{1}$, resulting in the Bell state $\frac{1}{\sqrt{2}}\left(|00\rangle+|11\rangle\right)_{12}\otimes|0\cdots00\rangle$
\cite{three-GHZ2,Bell-state2}. Following the above described method,
we can generate large even-number GHZ states. 

\section{preparation errors and fidelity }

In this section, we analyze the experimental feasibility of our scheme
by investigating preparation errors. Errors mainly come from two aspects.
The first one is the precision of the control of the interaction time.
For each ``rotation'' operation, if the interaction time is not
exactly $T$, the state realized is not a exact GHZ state. The second
is the environmental noises. We refer to recent experimental achievement
to discuss the feasibility of our scheme. The operations of switching
on (off) and dynamically modulating of inter-qubit coupling in circuit
QED can be in nanosecond timescales \cite{Martinis-PRL-2014}. The
typical qubit-qubit coupling strength is several megahertz \cite{naturePhysics_Martinis},
so the characteristic time $T$ for excitation circulation is hundreds
of nanosecond. The switching time is much less than $T$, therefore,
the precision of the control of the interaction time made the error
acceptable.

\begin{figure}
\begin{centering}
\includegraphics[width=9cm]{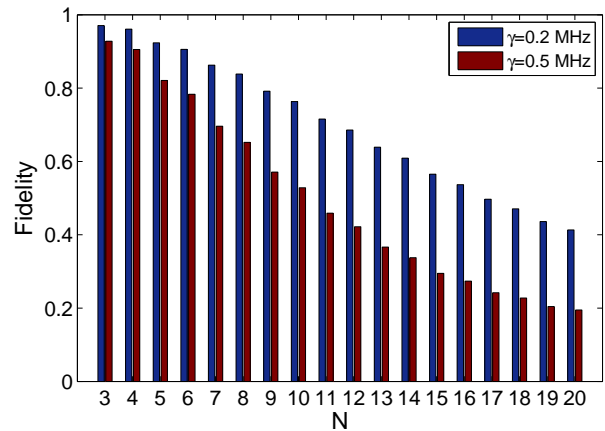}
\par\end{centering}
\caption{(Color online) Fidelity for different \textbf{\textit{N}}-qubit GHZ
states. Two kinds of bars corresponding to decay rate of qubit $\gamma=0.2\thinspace\mathrm{MHz}$
(blue bar) and $\gamma=0.5\thinspace\mathrm{MHz}$ (red bar). The
characteristic time for excitation circulation $T=100\thinspace\mathrm{ns}$.
\label{fig:fidelity-for-different}}
\end{figure}

The qubit relaxation and dephasing during the operation certainly
changes the final output state. In contrast to natural atoms, an advantage
of superconducting qubit is that its coherence time can be large as
tens of microseconds \cite{T1-1,T1-2,T1-3}. We calculate the fidelity
for \textit{N}-qubit GHZ state in our scheme. To be on the safe said,
we assume for the qubit $T=100\thinspace\mathbf{\mathrm{ns}}$, and
choose two decay rates $\gamma=0.2\thinspace\mathrm{MHz}$ and $\gamma=0.5\thinspace\mathrm{MHz}$
for reference. The results are shown in Fig. \ref{fig:fidelity-for-different},
where for even-number GHZ state, we neglect the errors in preparing
initial Bell state. 

Definitely, there will be other complicated factors influencing the
fidelity of GHZ state in real preparation process. Here, our analysis
is optimistic but the parameter we chosen for qubit decay rate is
not rigorous compare to existing experiments. In the Ref. \cite{naturePhysics_Martinis},
a similar circulation of photons can exist several cycles in the coherence
time of superconducting qubit. Therefore, we have confidence that
our scheme can work in appropriate circuit QED system.

\section{conclusion}

In conclusion, we propose a totally new mechanism for generation of
multiqubit GHZ states. The essential of this scheme is the directional
excitation circulation in three-qubit coupling loop, and this phenomenon
is derived from the complex hopping amplitude which break the time-reversal
symmetry. We can realize the pure imaginary hopping constant by dynamically
modulating the inter-qubit coupling which has been realized in circuit
QED experiment. There is another method to obtain the effective interaction
with imaginary hopping amplitude by dynamically modulating the frequency
of qubit \cite{Dawei2016PRL}. Based on these modulating methods,
our scheme for generation of multiqubit GHZ state can be applied to
other system. In this paper, we employ superconducting qubit to demonstrate
the experimental feasibility of our scheme. Our results of fidelity
analysis show the scheme is realizable with the existing experimental
conditions.
\begin{acknowledgments}
W.F. would like to acknowledge support from the China Scholarship
Council (Grant No. 201504890006). D-W.W. and H.C. would like to acknowledge
support from the Office of Naval Research (Award No. N00014-16-1-3054)
and Robert A. Welch Foundation (Grant No. A-1261). 
\end{acknowledgments}

\end{document}